\documentclass[]{amsart}

\usepackage{amsmath} 
\usepackage{bbm}
\usepackage{amsmath} 
\usepackage{amsfonts}
\usepackage{amssymb}
\usepackage{amsthm}
\usepackage{graphicx}
\usepackage{cite}

\usepackage{rcs}
\RCS $Revision$
\RCS $Date$
\newsavebox{\versioninfo}
\begin{lrbox}{\versioninfo}
\mbox{\small \RCSDate}
\end{lrbox}

\newcommand{\st}{\mathop{\hbox{\rm subject to}\;}}

\newcommand{\norm}[1]{\|#1\|}

\newcommand{\abs}[1]{\left|#1\right|}

\newcommand{\Real}{\mathbb{R}}

\renewcommand{\text}[1]{\quad\mbox{#1}\quad}  

\newcommand{\R}{{\mathbb{R}}}

\newcommand{\Ref}[1] {(\ref{#1})}

\newcommand{\de}{\delta}

\newcommand{\ddt}{\frac{d}{dt}}

\newtheorem{theorem}{Theorem}[section]

\newtheorem{definition}[theorem]{Definition}

\newtheorem{proposition}[theorem]{Proposition}

\begin{document}
\begin{center} \Huge{Functional Bregman Divergence and \\ Bayesian
Estimation of Distributions\\}\end{center}

\normalsize  \noindent B. A. Frigyik,
  Purdue University,
  bfrigyik@math.purdue.edu\\
  S. Srivastava, Univ. of Washington, santosh@amath.washington.edu\\
  M. R. Gupta, Univ. of Washington, gupta@ee.washington.edu

\section*{Abstract}
\emph{A class of distortions termed functional Bregman divergences
is defined, which includes squared error and relative entropy. A
functional Bregman divergence acts on functions or distributions,
and generalizes the standard Bregman divergence for vectors and a
previous pointwise Bregman divergence that was defined for
functions. A recently published result showed that the mean
minimizes the expected Bregman divergence. The new functional
definition enables the extension of this result to the continuous
case to show that the mean minimizes the expected functional Bregman
divergence over a set of functions or distributions. It is shown how
this theorem applies to the Bayesian estimation of distributions.
Estimation of the uniform distribution from independent and
identically drawn samples is used as a case study.}

\section{Overview}
Bregman divergences are a useful set of distortion functions that
include squared error, relative entropy, logistic loss, Mahalanobis
distance, and the Itakura-Saito function. Bregman divergences are
popular in statistical estimation and information theory. Analysis
using the concept of Bregman divergences has played a key role in
recent advances in statistical learning \cite{Jordan:06, Murata:04,
Singer:02, Kivinen:01, Lafferty:99, SrivastavaGupta:06, JonesByrne,
knnLaplace, BDA7}, clustering \cite{Ghosh:05, Nock:06}, inverse
problems \cite{Besnerais:99}, maximum entropy estimation
\cite{Altun:06}, and the applicability of the data processing
theorem \cite{Pardo:97}. Recently, it was discovered that the mean
is the minimizer of the expected Bregman divergence for a set of
$d$-dimensional points \cite{Banerjee:05, Ghosh:05}.

In this paper we define a functional Bregman divergence that applies
to functions and distributions, and we show that this new definition
is equivalent to Bregman divergence applied to vectors. The
functional definition generalizes a pointwise Bregman divergence
that has been previously defined for measurable functions
\cite{JonesByrne, Csiszar:95}, and thus extends the class of
distortion functions that are Bregman divergences; see Section
\ref{sec:biasExample} for an example. Most importantly, the
functional definition enables one to solve functional minimization
problems using standard methods from the calculus of variations; we
extend the recent result on the expectation of vector Bregman
divergence \cite{Banerjee:05, Ghosh:05} to show that the mean
minimizes the expected Bregman divergence for a set of functions or
distributions. We show how this theorem links to Bayesian estimation
of distributions. For distributions from the exponential family
distributions, many popular divergences, such as relative entropy,
can be expressed as a (different) Bregman divergence on the
exponential distribution parameters. The functional Bregman
definition enables stronger results and a more general application.
\\

In Section 1 we state a functional definition of the Bregman
divergence and give examples for total squared difference, relative
entropy, and squared bias. The relationship between the functional
definition and previous Bregman definitions is established. In
Section 2 we present the main theorem: that the expectation of a set
of functions minimizes the expected Bregman divergence. In Section 3
we discuss the role of this theorem in Bayesian estimation, and as a
case study compare different estimates for the uniform distribution
given independent and identically drawn samples.  For ease of
reference, Appendix A contains relevant definitions and results from
functional analysis and the calculus of variations. In Appendix B we
show that the functional Bregman divergence has many of the same
properties as the standard vector Bregman divergence. Proofs are in
Appendix C.

\section{Functional Bregman Divergence}
Let $\left(\Real^d,\Omega,\nu\right)$ be a measure space, where
$\nu$ is a Borel measure, $d$ is a positive integer, and define a
set of functions $\mathcal{A} = \left\{a\in L^p(\nu)\st a:\Real^d \to
 \Real,\;  a\geq 0 \right\}$ where $1 \leq p \leq \infty$.

\begin{definition}[Functional Definition of Bregman Divergence]
Let $\phi:L^{p}(\nu)\rightarrow \Real$ be a strictly convex,
twice-continuously Fr\'{e}chet-differentiable functional. The
Bregman divergence $d_{\phi}: \mathcal{A}\times \mathcal{A}
\rightarrow [0,\infty)$  is defined for all $f,g\in\mathcal{A}$ as
\begin{equation}\label{eqn:BregmanDef}
d_{\phi}[f,g] = \phi[f] - \phi[g] - \delta \phi[g;f-g],
\end{equation}
where $\delta \phi[g; \cdot]$ is the Fr\'{e}chet derivative of
$\phi$ at $g$.
\end{definition}
Here, we have used the Fr\'{e}chet derivative, but the definition
(and results in this paper) can be easily extended using more
general definitions of derivatives; a sample extension is given in
Section \ref{sec:relent}.

The functional Bregman divergence has many of the same properties as
the standard vector Bregman divergence, including non-negativity,
convexity, linearity, equivalence classes, linear separation, dual
divergences, and a generalized Pythagorean inequality. These
properties are established in Appendix B.

\subsection{Examples}\label{sec:examples}
Different choices of the functional $\phi$ lead to different Bregman
divergences.  Illustrative examples are given for squared error,
squared bias, and relative entropy. Functionals for other Bregman
divergences can be derived based on these examples, from the example
functions for the discrete case given in Table 1 of
\cite{Banerjee:05}, and from the fact that $\phi$ is a strictly
convex functional if it has the form $\phi(g)=\int
\tilde\phi(g(t))dt$
where $\tilde\phi:\Real \rightarrow \Real$, $\tilde\phi$ is
strictly convex and $g$ is in some well-defined vector space of
functions \cite{Rockafellar:68}.

\subsubsection{Total Squared Difference}
Let $\phi[g]=\int g^2 d\nu$, where $\phi:L^2(\nu)\to \Real$, and let
$g, f, a \in L^2(\nu)$. Then
\begin{eqnarray*}
  \phi[g+a]-\phi[g] &=&\int (g+a)^2d\nu-\int g^2d\nu\\
  &=&2\int ga d\nu+\int a^2 d\nu.
\end{eqnarray*}
Because
\begin{equation*}
  \frac{\int a^2 d\nu}{\norm{a}_{L^2(\nu)}}=
  \frac{\norm{a}_{L^2(\nu)}^2}{\norm{a}_{L^2(\nu)}}=\norm{a}_{L^2(\nu)}\to
  0
\end{equation*}
as $a\to 0$ in $L^2(\nu)$,
\begin{equation*}
\de\phi[g;a]=2\int ga d\nu,
\end{equation*}
which is a continuous linear functional in $a$. Then, by definition of the
second Fr\'{e}chet derivative,
\begin{eqnarray*}
  \de^2\phi[g;b,a] &=&\de\phi[g+b;a]-\de\phi[g;a] \\
  &=& 2\int (g+b)a d\nu-2\int ga d\nu\\
  &=& 2\int ba d\nu.
\end{eqnarray*}
Thus  $\de^2\phi[g;b,a]$ is a quadratic form, where $\de^2\phi$ is
actually independent of $g$ and strongly positive since
\begin{equation*}
  \de^2\phi[g;a,a]=2\int a^2 d\nu= 2 \norm{a}_{L^2(\nu)}^2
\end{equation*}
for all $a\in L^2(\nu)$, which implies that $\phi$ is strictly convex
and
\begin{eqnarray*}
d_\phi[f,g] &=&\int f^2 d\nu-\int g^2 d\nu- 2\int g(f-g) d\nu\\
&=& \int (f-g)^2 d\nu \\
&=&\norm{f-g}_{L^2(\nu)}^2.
\end{eqnarray*}

\subsubsection{Squared Bias}\label{sec:biasExample}
Under definition (\ref{eqn:BregmanDef}), squared bias is a Bregman
divergence, this we have not previously seen noted in the literature
despite the importance of minimizing bias in estimation \cite{HTF}.

Let $\phi[g]=\left( \int g d\nu\right)^2$, where $\phi:
L^1(\nu)\to\Real$. In this case
\begin{eqnarray}
\phi[g+a]-\phi[g]&=&\left(\int g d\nu +\int a
d\nu\right)^2-\left(\int g
    d\nu\right)^2 \nonumber\\
  &=&2\int g d\nu\int a d\nu +\left(\int a d\nu\right)^2.
  \label{eqn:number}
\end{eqnarray}
Note that $2\int g d\nu\int a d\nu$ is a continuous linear
functional on $L^1(\nu)$ and $\left(\int a
d\nu\right)^2\leq\norm{a}_{L^1(\nu)}^2$, so that
\begin{equation*}
  0\leq\frac{\left(\int a d\nu\right)^2}{\norm{a}_{L^1(\nu)}}\leq
  \frac{\norm{a}_{L^1(\nu)}^2}{\norm{a}_{L^1(\nu)}}=\norm{a}_{L^1(\nu)}.
\end{equation*}
Thus from (\ref{eqn:number}) and the definition of the Fr\'{e}chet
derivative,
\begin{equation*}
\de\phi[g;a]=2\int g d\nu\int a d\nu.
\end{equation*}
By the definition of the second Fr\'{e}chet derivative,
\begin{eqnarray*}
\de^2\phi[g;b,a] &=& \de\phi[g+b;a]-\de\phi[g;a] \\
&=& 2\int (g+b) d\nu \int a d\nu-2\int g d\nu\int a d\nu \\
&=& 2\int b d\nu \int a d\nu
\end{eqnarray*}
is another quadratic form, and $\de^2\phi$ is independent of $g$.

Because the functions in $\mathcal{A}$ are positive, $\de^2\phi$ is
strongly positive on $\mathcal{A}$ (which again implies that $\phi$ is
strictly convex):
\begin{equation*}
  \de^2\phi[g;a,a]=2\left(\int a d\nu \right)^2= 2 \norm{a}_{L^1(\nu)}^2 \geq 0
\end{equation*}
for
$a\in\mathcal{A}$. The Bregman divergence is thus
\begin{eqnarray*}
\lefteqn{d_\phi[f,g]} \\
 &=&\left(\int f d\nu\right)^2- \left(\int g
d\nu\right)^2-
2\int g d\nu\int (f-g) d\nu\\
 &=& \left(\int f d\nu\right)^2 +\left(\int g
  d\nu\right)^2 -2\int g d\nu\int f d\nu\\
  &=&\left(\int (f-g)
  d\nu\right)^2\\
  &\leq& \norm{f-g}_{L^1(\nu)}^2.
\end{eqnarray*}

\subsubsection{Relative Entropy of Simple Functions}\label{sec:relent}
Let $(X,\Sigma,\nu)$ be a measure space. We denote by $\mathcal{S}$
the collection of all measurable simple functions on
$(X,\Sigma,\nu)$, that is, the set of functions which can be written
as a finite linear combination of indicator functions. If
$g\in\mathcal{S}$ then it can be expressed as
\begin{equation*}
  g(x)=\sum_{i=0}^t\alpha_i I_{T_i};\quad \alpha_0=0,
\end{equation*}
where $I_{T_i}$ is the indicator function of the set $T_i$ and
$\{T_i\}_{i=0}^t$ is a collection of mutually disjoint measurable
sets with the property that $X=\bigcup_{i=0}^t T_i$. We adopt the
convention, that $T_0$ is the set on which $g$ is zero and therefore
$\alpha_i\neq 0$ if $i\neq 0$. The set
$\left(\mathcal{S},\norm{\cdot}_{L^\infty(\nu)}\right)$ is a normed
vector space. In this case
\begin{equation}\label{eqn:hippo}
  \int_X g\ln g d\nu=\sum_{i=1}^t \int_{T_i}\alpha_i\ln \alpha_i
  d\nu,
\end{equation}
since $0\ln 0=0$.


Note that the integral in (\ref{eqn:hippo}) exists and is finite for
$g \in \mathcal{S}$ if $g \in L^1(\nu)$ and $g \geq 0$. This implies
that $\nu(T_i)<\infty$ for all $1\leq i\leq t$, while the measure of
$T_0$ could be infinity. For this reason, consider the normed vector
space $(L^1(\nu)\cap\mathcal{S},$$\norm{\cdot}_{L^\infty(\nu)})$,
where $(L^1(\nu)\cap\mathcal{S})$$\subset \mathcal{S}$$\subset
L^\infty(\nu)$. Let $\mathcal{W}$ be the set (not necessarily a
vector space) of functions satisfying the conditions mentioned above
-- that is, let
\begin{equation*}
  \mathcal{W}=\{g\in L^1(\nu)\cap\mathcal{S} \st g\geq 0 \}.
\end{equation*}
Define the functional $\phi$ on $\mathcal{W}$,
\begin{equation}\label{eqn:moose}
  \phi[g]=\int_X g\ln g\,d\nu, \quad g \in \mathcal{W}.
\end{equation}
The functional $\phi$ is not Fr\'{e}chet-differentiable at $g$ because
in general it cannot be guaranteed that $g+h$ is non-negative for all
functions $h$ in the underlying normed vector space
$\left(L^1(\nu)\cap\mathcal{S},\norm{\cdot}_{L^\infty(\nu)}\right)$
with norm smaller than any prescribed $\epsilon>0$. However, a
generalized G\^ateaux derivative can be defined if we limit the
perturbing function $h$ to a vector subspace.

Let $\mathcal{G}$ be the subspace of
$\left(L^1(\nu)\cap\mathcal{S},\norm{\cdot}_{L^\infty(\nu)}\right)$
defined by
\begin{equation*}
  \mathcal{G}=\{f\in L^1(\nu)\cap\mathcal{S} \st f\,d\nu\ll
  g\,d\nu\}.
\end{equation*}

It is straightforward to show that $\mathcal{G}$ is vector space.
We define the generalized G\^ateaux derivative of $\phi$ at $g\in
\mathcal{W}$ to be the linear operator $\delta_G\phi[g;\cdot]$ if
\begin{equation}\label{eqn:antler}
  \lim_{\substack{\norm{h}_{L^\infty(\nu)}\to 0\\ h\in\mathcal{G}}}
  \frac{\abs{\phi[g+h]-\phi[g]-\delta_G\phi[g;h]}}
    {\norm{h}_{L^\infty(\nu)}}=0.
\end{equation}
Note, that $\delta_G\phi[g;\cdot]$ is not linear in general, but it
is on the vector space $\mathcal{G}$. In general, if $\mathcal{G}$
is the entire underlying vector space then (\ref{eqn:antler}) is the
Fr\'echet derivative, and if $\mathcal{G}$ is the span of only one
element from the underlying vector space then (\ref{eqn:antler}) is
the G\^ateaux derivative. Here, we have generalized the G\^ateaux
derivative for the present case that $\mathcal{G}$ is a subspace of
the underlying vector space.

It remains to be shown that given the functional (\ref{eqn:moose}),
the derivative (\ref{eqn:antler}) exists and yields relative
entropy. Consider the solution
\begin{equation}\label{eqn:kiwi}
\delta_G\phi[g;h]=\int_X (1+\ln g)hd\nu,
\end{equation}
which coupled with (\ref{eqn:moose}) does yield relative entropy. We
complete the proof by showing that (\ref{eqn:kiwi}) satisfies
(\ref{eqn:antler}). Note that
\begin{align}
\phi[g+h]-\phi[g]-\delta_G\phi[g;h]
  &=\int_X (h+g)\ln\frac{h+g}{g}-h d\nu \nonumber \\
  &=\int_E(h+g)\ln\frac{h+g}{g}-h d\nu,\label{eqn:cat}
\end{align}
where $E$ is the set on which $g$ is not zero.

Because $g\in\mathcal{W}$, there are $m,M>0$ such that $m\leq g\leq
M$ on $E$. Let $h\in\mathcal{G}$ be such that
$\norm{h}_{L^\infty(\nu)}\leq m$, then $g+h\geq 0$. Our goal is to
find a lower and an upper bound for the expression
\begin{equation*}
  \frac{\phi[g+h]-\phi[g]-\delta_G\phi[g;h]} {\norm{h}_{L^\infty(\nu)}}
\end{equation*}
such that both bounds go to $0$ as $\norm{h}_{L^\infty(\nu)}\to 0$.
We start with bounding the integrand from above:
\begin{equation*}
  (h+g)\ln\frac{h+g}{g}-h\leq
  (h+g)\frac{h}{g}-h=\frac{h^2}{g},
\end{equation*}
and therefore
\begin{eqnarray*}
\frac{\phi[g+h]-\phi[g]-\delta_G\phi[g;h]}
    {\norm{h}_{L^\infty(\nu)}}&\leq&\frac{1}{\norm{h}_{L^\infty(\nu)}}\int_E \frac{h^2}{g}d\nu \\
    &\leq& \frac{1}{m}\int_E \abs{h} d\nu \\
    &\leq& \frac{1}{m}\norm{h}_{L^{1}(\nu)}.
\end{eqnarray*}

We can use Jensen's inequality to find a lower bound for the
integral \Ref{eqn:cat}. In order to use the inequality we have to
rewrite the equation. We begin with the first term of the integrand,
\begin{eqnarray*}
\lefteqn{\int_E(h+g)\ln\frac{h+g}{g}d\nu}\\
&=&  \int_E\frac{h+g}{g}\left(\ln\frac{h+g}{g}\right)gd\nu,\\
&=&  \norm{g}_{L^1(\nu)} \int_E\frac{h+g}{g}\ln\frac{h+g}{g}\frac{g}
  {\norm{g}_{L^1(\nu)}}d\nu\\
  &=&\norm{g}_{L^1(\nu)} \int_E\lambda\left(\frac{h+g}{g}\right)
  d\tilde{\nu},
\end{eqnarray*}
where the measure $d\tilde{\nu}=\frac{g}{\norm{g}_{L^1(\nu)}}d\nu$
is a probability measure and $\lambda(x)=x\ln x$ is a convex
function on $(0,\infty)$. Let $M_0 = \norm{g}_{L^1(\nu)}$. By
Jensen's inequality
\begin{eqnarray*}
  \lefteqn{M_0\int_E\lambda\left(\frac{h+g}{g}\right) d\tilde{\nu}} \\
  &\geq&
  M_0\lambda\left( \int_E \frac{h+g}{g} d\tilde{\nu}\right)\\
  &=&M_0\lambda\left(\int_E \frac{h}{M_0} d\nu+ \int_E  d\tilde{\nu}
  \right)\\
  &=&M_0\lambda\left(\frac{1}{M_0}\int_E h\,d\nu+1\right)\\
  &=&\left(\int_E h\,d\nu+M_0\right)\ln\left( \frac{1}{M_0}\int_E h\,d\nu+1
  \right).
\end{eqnarray*}
Thus we can bound the integral in \Ref{eqn:cat} from below:
\begin{eqnarray*}
\lefteqn{\int_E(h+g)\ln\frac{h+g}{g}-h d\nu}\\
  &\geq&\left(\int_E h\,d\nu+M_0\right)\ln\left( \frac{1}{M_0}\int_E h\,d\nu+1
  \right)-\int_E h\,d\nu\\
  &=&\int_E h\,d\nu\ln \left( \frac{1}{M_0}\int_E h\,d\nu+1\right)\\
  &&+M_0\ln \left( \frac{1}{M_0}\int_E h\,d\nu+1\right)-\int_E h\,d\nu.
\end{eqnarray*}
If $\int_E h\,d\nu=0$, then the integral in \Ref{eqn:cat} is
non-negative. The more interesting case is when $\int_E h\,d\nu\neq
0$. Then,
\begin{eqnarray*}
\lefteqn{\frac{\phi[g+h]-\phi[g]-\delta_G\phi[g;h]}
  {\norm{h}_{L^\infty(\nu)}}} \\
  &\geq&\frac{\int_E h\,d\nu}{\norm{h}_{L^\infty(\nu)}}
  \ln \left( \frac{1}{M_0}\int_E h\,d\nu+1\right)\\
  &&+\frac{M_0}{\norm{h}_{L^\infty(\nu)}}
  \ln \left( \frac{1}{M_0}\int_E h\,d\nu+1\right)-
  \frac{\int_E h\,d\nu}{\norm{h}_{L^\infty(\nu)}}\\
  &\geq&\frac{\int_E h\,d\nu}{\norm{h}_{L^\infty(\nu)}}
  \ln \left( \frac{1}{M_0}\int_E h\,d\nu+1\right)\\
  &&+\left[\frac{M_0\ln \left( \frac{1}{M_0}\int_E h\,d\nu+1\right)}
  {\int_E h\,d\nu}-1\right] \frac{\int_E h\,d\nu}
  {\norm{h}_{L^\infty(\nu)}}.
\end{eqnarray*}
As $\int_E h\,d\nu\to 0$,
\begin{gather*}
  \ln \left( \frac{1}{M_0}\int_E h\,d\nu+1\right) \to 0,
\end{gather*}
and
\begin{gather*}
  \frac{M_0\ln \left( \frac{1}{M_0}\int_E h\,d\nu+1\right)}
  {\int_E h\,d\nu}-1\to 0.
\end{gather*}
We finish the proof by showing that there is a constant $K$ which is
independent of $h$ such that
\begin{equation}\label{eqn:tahiti}
  \abs{\int_E h\,d\nu}\leq \norm{h}_{L^1(\nu)}\leq K
  \norm{h}_{L^\infty(\nu)}.
\end{equation}
If (\ref{eqn:tahiti}) is shown, then $\int_E h\,d\nu\to 0$ and
$\norm{h}_{L^1(\nu)}\to 0$ as $\norm{h}_{L^\infty(\nu)}\to 0$, and
coupling those relationships with the fact that
\begin{equation*}
  \frac{\abs{\int_E h\,d\nu}}{\norm{h}_{L^\infty(\nu)}}\leq K
\end{equation*}
establishes (\ref{eqn:antler}). Because $h\in\mathcal{G}$, $h$ can
be expressed as
\begin{equation*}
  h=\sum_{i=0}^{v} \beta_i I_{V_i};\quad \beta_0=0,
\end{equation*}
where $\{V_i\}_{i=0}^{v}$ is a collection of mutually disjoint
measurable sets with the property that $X =\bigcup_{i=0}^v V_i$.
Also, because $h\,d\nu\ll g\,d\nu$, there is a set $N(h)$ such that
$\nu(N(h))=0$ and
\begin{equation*}
  \bigcup_{i=1}^v V_i\subset \left(\bigcup_{i=1}^t T_i \cup N(h)\right).
\end{equation*}
This implies that there is a $K$ independent of $h$ such that
\begin{equation*}
  \sum_{i=1}^v\nu(V_i)\leq \sum_{i=1}^t\nu(T_i)=K.
\end{equation*}
Finally,
\begin{eqnarray*}
\int \abs{h}d\nu&=&\sum_{i=1}^v\abs{\beta_i}\nu(V_i)\\
 &\leq&
  \norm{h}_{L^\infty(\nu)} \sum_{i=1}^v\nu(V_i)\\
  &\leq&
  \norm{h}_{L^\infty(\nu)}K.
\end{eqnarray*}

\subsection{Relationship to Other Bregman Divergence Definitions}
Two propositions establish the relationship of the functional
Bregman divergence to other Bregman divergence definitions.
\begin{proposition}[Functional Bregman Divergence Generalizes Vector Bregman
Divergence] The functional definition (\ref{eqn:BregmanDef}) is a
generalization of the standard vector Bregman divergence
\begin{equation}\label{eqn:standard}
d_{\tilde\phi}(x,y) = \tilde\phi(x) - \tilde\phi(y) - \nabla
\tilde\phi(y)^T(x-y),
\end{equation}
where $x,y \in \Real^n$, and $\tilde\phi:\Real^n\to\Real$ is
strictly convex and twice differentiable.
\end{proposition}

Jones and Byrne describe a general class of divergences between
functions using a pointwise formulation \cite{JonesByrne}.
Csisz\'{a}r specialized the pointwise formulation to a class of
divergences he termed \emph{Bregman distances} $B_{s, \nu}$
\cite{Csiszar:95}, where given a $\sigma$-finite measure space $(X,
\Omega, \nu)$, and non-negative measurable functions $f(x)$ and
$g(x)$, $B_{s, \nu}(f,g)$ equals
\begin{equation}\label{eqn:pointwise}
\int s(f(x)) - s(g(x)) - s'(g(x))(f(x)-g(x)) d\nu(x).
\end{equation}
The function $s: (0, \infty) \rightarrow \Real$ is constrained to be
differentiable and strictly convex, and the limit $\lim_{x
\rightarrow 0} s(x)$ and $\lim_{x \rightarrow 0} s'(x)$ must exist,
but not necessarily finite. The function $s$ plays a role similar to
the function $\phi$ in the functional Bregman divergence; however,
$s$ acts on the range of the functions $f,g$, whereas $\phi$ acts on
the pair of functions $f,g$.

\begin{proposition}[Functional Definition Generalizes Pointwise Definition]
Given a pointwise Bregman divergence as per (\ref{eqn:pointwise}),
an equivalent functional Bregman divergence can be defined as per
(\ref{eqn:BregmanDef}) if the measure $\nu$ is finite. However,
given a functional Bregman divergence $d_{\phi}[f,g]$, there is not
necessarily an equivalent pointwise Bregman divergence.
\end{proposition}

\section{Minimum Expected Bregman Divergence}
Consider two sets of functions (or distributions), $\mathcal{M}$ and
$\mathcal{A}$. Let $F\in \mathcal{M}$ be a random function with
realization $f$. Suppose there exists a probability distribution
$P_F$ over the set $\mathcal{M}$, such that $P_F(f)$ is the
probability of $f \in \mathcal{M}$. For example, consider the set of
Gaussian distributions, and given samples drawn independently and
identically from a randomly selected Gaussian distribution $N$, the
data imply a posterior probability $P_N(\mathcal{N})$ for each
possible generating realization of a Gaussian distribution
$\mathcal{N}$. The goal is to find the function $g^* \in
\mathcal{A}$ that minimizes the expected Bregman divergence between
the random function $F$ and any function $g \in \mathcal{A}$. The
following theorem shows that if the set of possible minimizers
$\mathcal{A}$ includes $E_{P_F}[F]$, then $g^* = E_{P_F}[F]$
minimizes the expectation of any Bregman divergence.

The theorem applies only to a set of functions $\mathcal{M}$ that
lie on a finite-dimensional manifold $M$ for which a differential
element $dM$ can be defined. For example, the set $\mathcal{M}$
could be parameterized by a finite number of parameters, or could be
a set of functions that can be decomposed into a finite set of $d$
basis functions $\{\psi_1, \psi_2, \ldots, \psi_d \}$ such that each
$f$ can be expressed as
\begin{equation*}
f = \sum_{j=1}^d c_j \psi_j,
\end{equation*}
where $c_j \in \Real$
for all $j$. The theorem
requires slightly stronger conditions on $\phi$ than the definition
of the Bregman divergence (\ref{eqn:BregmanDef}) requires.

\begin{theorem}[Minimizer of the Expected Bregman Divergence]
Let $\delta^2\phi[f;a,a]$ be a strongly positive quadratic form, and
let $ \phi \in \mathcal{C}^3(L^{1}(\nu);\Real)$ be a three-times
continuously Fr\'echet-differentiable functional on $L^1(\nu)$. Let
$\mathcal{M}$ be a set of functions that lie on a finite-dimensional
manifold $M$, and have associated differential element $dM$. Suppose
there is a probability distribution $P_F$ defined over the set
$\mathcal{M}$. Suppose the function $g^*$ minimizes the expected
Bregman divergence between the random function $F$ and any function $g
\in \mathcal{A}$ such that
\begin{equation*} g^* =
\arg \inf_{g \in \mathcal{A}} E_{P_F}[d_{\phi}(F,g)].
\end{equation*}
Then, if $g^*$ exists, it is given by
\begin{equation}\label{eqn:FunctionalMean}
g^* = \int_{M} f P(f) dM = E_{P_F}[F].
\end{equation}
\end{theorem}

\section{Bayesian Estimation}\label{sec:bayesian}
Theorem II.1 can be applied to a set of distributions to find the
Bayesian estimate of a distribution given a posterior or likelihood.
For parametric distributions parameterized by $\theta \in \Real^n$,
a probability measure $\Lambda(\theta)$, and some risk function
$R(\theta, \psi)$, $\psi \in \Real^n$,  the Bayes estimator is
defined \cite{Lehmann:98} as
\begin{equation}\label{eqn:handel}
\hat{\theta} = \arg \inf_{\psi \in \Real^n} \int R(\theta, \psi)
d\Lambda(\theta).
\end{equation}
That is, the Bayes estimator minimizes some expected risk in terms
of the parameters. It follows from recent results \cite{Banerjee:05}
that $\hat{\theta} = E[\Theta]$ if the risk $R$ is a Bregman
divergence, where $\Theta$ is the random variable whose realization
is $\theta$.

The principle of Bayesian estimation can be applied to the
distributions themselves rather than to the parameters:
\begin{equation}\label{eqn:stravinsky}
\hat{g} = \arg \inf_{g \in \mathcal{A}}  \int_{M} R(f, g) P_F(f) dM,
\end{equation}
where $P_F(f)$ is a probability measure on the distributions $f \in
\mathcal{M}$, $dM$ is a differential element for the
finite-dimensional manifold $M$, and $\mathcal{A}$ is either the
space of all distributions or a subset of the space of all
distributions, such as the set $\mathcal{M}$. When the set
$\mathcal{A}$ includes the distribution $E_{P_F}[F]$ and the risk
function $R$ in (\ref{eqn:stravinsky}) is a Bregman divergence, then
Theorem II.1 establishes that $\hat{g}= E_{P_F}[F]$.

For example, in recent work, two of the authors derived the mean
class posterior distribution for each class for a Bayesian quadratic
discriminant analysis classifier \cite{SrivastavaGupta:06}, and
showed that the classification results were superior to
parameter-based Bayesian quadratic discriminant analysis.

Of particular interest for estimation problems are the Bregman
divergence examples given in Section \ref{sec:examples}: total squared
difference (mean squared error) is a popular risk function in regression \cite{HTF};
minimizing relative entropy leads to useful theorems for large
deviations and other statistical subfields \cite{CoverThomas}; and
analyzing bias is a common approach to characterizing and
understanding statistical learning algorithms \cite{HTF}.

\subsection{Case Study: Estimating a Scaled Uniform Distribution}
As an illustration, we present and compare different estimates of a
scaled uniform distribution given independent and identically drawn
samples. Let the set of uniform distributions over $[0, \theta]$ for
$\theta \in \Real^+$ be denoted by $\mathcal{U}$. Given independent
and identically distributed samples $X_1, X_2, \ldots, X_n$ drawn
from an unknown uniform distribution $f \in \mathcal{U}$, the
generating distribution is to be estimated. The risk function $R$ is
taken to be squared error or total squared error depending on
context.

\subsubsection{Bayesian Parameter Estimate}\label{sec:1}
Depending on the choice of the probability measure
$\Lambda(\theta)$, the integral (\ref{eqn:handel}) may not be
finite; for example, using the likelihood of $\theta$ with Lebesgue
measure the integral is not finite. A standard solution is to use a
gamma prior on $\theta$ and Lebesgue measure. Let $\Theta$ be a
random parameter with realization $\theta$, let the gamma
distribution have parameters $t_1$ and $t_2$, and denote the maximum
of the data as $X_{\max} = \max \{X_1, X_2, \ldots, X_n\}$. Then a
Bayesian estimate is formulated \cite[p. 240, 285]{Lehmann:98}:
\begin{eqnarray}\label{eqn:topEstimate1}
\lefteqn{E[\Theta|\{X_1, X_2, \ldots, X_n\}, t_1, t_2]}\nonumber\\
& =& \frac{\int_{X_{\max}}^{\infty}
\theta\frac{1}{\theta^{n+t_1+1}}e^{\frac{-1}{\theta t_2}} d\theta}
{\int_{X_{\max}}^{\infty}
\frac{1}{\theta^{n+t_1+1}}e^{\frac{-1}{\theta t_2}} d\theta}.
\end{eqnarray}
The integrals can be expressed in terms of the chi-squared random
variable $I_v^2$ with $v$ degrees of freedom:
\begin{eqnarray}\label{eqn:topEstimate}
\lefteqn{E[\Theta|\{X_1, X_2, \ldots, X_n\}, t_1, t_2] =}&& \nonumber\\
&&\frac{1}{t_2(n+t_1-a)}\frac{P(\chi^2_{2(n+t_1-1)} <
\frac{2}{t_2X_{\max}})}{P(\chi^2_{2(n+t_1)} <
\frac{2}{t_2X_{\max}})}.
\end{eqnarray}
Note that (\ref{eqn:handel}) presupposes that the best solution is
also a uniform distribution.

\subsubsection{Bayesian Uniform Distribution Estimate}\label{sec:2} If one restricts
the minimizer of (\ref{eqn:stravinsky}) to be a uniform
distribution, then (\ref{eqn:stravinsky}) is solved with
$\mathcal{A} = \mathcal{U}$. Because the set of uniform distributions does
not generally include its mean, Theorem II.1 does not apply, and
thus different Bregman divergences may give different minimizers for
(\ref{eqn:stravinsky}). Let $P_F$ be the likelihood of the data (no
prior is assumed over the set $\mathcal{U}$), and use the Fisher
information metric (\cite{Kass, Amari:00, Lebanon:05}) for $dM$. Then
the solution to (\ref{eqn:stravinsky}) is the uniform distribution
on $[0, 2^{1/n}X_{\max}]$. Using Lebesgue measure instead gives a
similar result: $[0, 2^{1/(n+1/2)}X_{\max}]$. We were unable to find
these estimates in the literature, and so their derivations are
presented in Appendix C.

\subsubsection{Unrestricted Bayesian Distribution Estimate}\label{sec:3}
When the only restriction placed on the minimizer $g$ in
(\ref{eqn:stravinsky}) is that $g$ be a distribution, then one can
apply Theorem II.1 and solve directly for the expected distribution
$E_{P_F}[F]$. Let $P_F$ be the likelihood of the data (no prior is
assumed over the set $\mathcal{U}$), and use the Fisher information
metric for $dM$. Solving (\ref{eqn:FunctionalMean}), noting that the
uniform probability of $x$ is $f(x) = 1/a$ if $x \leq a$ and zero
otherwise, and the likelihood of the $n$ drawn points is
$(1/X_{\max})^n$ if $a \geq X_{\max}$ and zero otherwise,
\begin{eqnarray}
g^*(x)&=&\frac{\int_{\max(x,X_{\max})}^{\infty}\left(\frac{1}{a}\right)
\left(\frac{1}{a^n}\right)\left(\frac{da}{a}\right)}
{\int_{X_{\max}}^{\infty} \frac{1}{a^n}\frac{da}{a}} \nonumber \\
&&\nonumber \\
&=&\frac{n\left(X_{\max}\right)^n}{(n+1)[\max(x,X_{\max})]^{n+1}}.\label{eqn:riteofspring}
\end{eqnarray}

\subsubsection{Projecting the Unrestricted Estimate onto the Set of
Uniform Distributions} Consider what happens when the unrestricted
solution $g^*(x)$  given in (\ref{eqn:riteofspring}) is projected
onto the set of uniform distributions with respect to squared error.
That is, we solve for the uniform distribution $h(x)$ over $[0, a]$
such that:
\begin{equation}\label{eqn:bestsub}
\hat{a} = \arg \min_{a \in [0,\infty)}
\int_{0}^{\infty}(h(x)-g^*(x))^2 dx.
\end{equation}
The problem is straightforward to solve using standard calculus and
yields the solution $\hat{a} = 2^{1/n}X_{\max}$. This is also the
solution to the problem (\ref{eqn:stravinsky}) when the minimizer is
restricted to be a uniform distribution and the Fisher information
metric over the uniform distributions is used (as discussed in Section \ref{sec:3}).
Thus, the projection of the unrestricted solution to (\ref{eqn:stravinsky}) onto the set
of uniform distributions is the same as the solution to
(\ref{eqn:stravinsky}) when the minimizer is restricted to be
uniform. We conjecture that under some conditions this property will
hold more generally: that the projection of the unrestricted
minimizer of (\ref{eqn:stravinsky}) onto the set $\mathcal{M}$ will
be equivalent to solving (\ref{eqn:stravinsky}) where the solution
is restricted to the set $\mathcal{M}$.


\subsection{Simulation} A simulation was done to compare the different
Bayesian estimators and the maximum likelihood estimator. The
simulation was run $1,000$ times; each time $n$ data points were
drawn independently and identically from the uniform over $[0, 1]$,
and estimates were formed. Figure 1 is a log-log plot of the average
squared errors between the estimated distribution and the true
distribution.

\begin{figure}
  \includegraphics[width=3.8in]{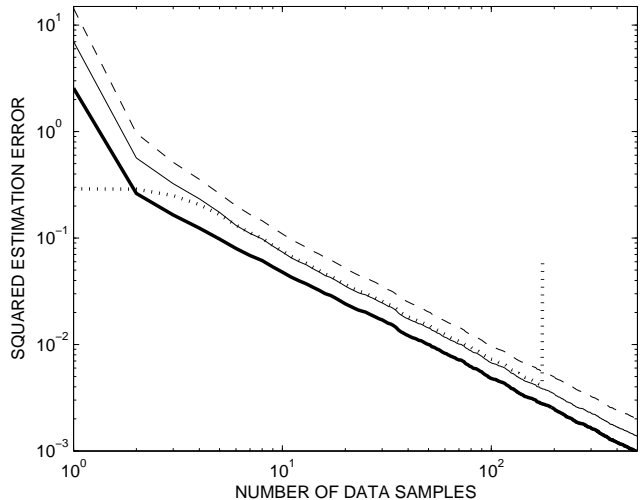}
  \caption{The plot shows the log of the squared error between an estimated
  distribution and a uniform $[0,1]$ distribution, averaged over one thousand
runs of the estimation simulation. The dashed line is the maximum
likelihood
  estimate,
  the dotted line is the Bayesian parameter estimate, the thick solid line is
  the Bayesian distribution
  estimate that solves (\ref{eqn:stravinsky}), and the thin solid line
  is the Bayesian distribution estimate that solves (\ref{eqn:stravinsky})
  when the minimizer is restricted to be uniform.}\label{fig:uniform1TenSamples}
\end{figure}

For the Bayesian parameter estimator given in
(\ref{eqn:topEstimate}), estimates were calculated for three
different sets of Gamma parameters, $(t_1 = 1, t_2 = 1)$, $(t_1 = 1,
t_2 = 3)$, and $(t_1=1, t_2=100)$. The plotted error is the minimum
of the three averaged errors for the different Gamma priors for each
$n$. The plotted Bayesian distribution estimates used the Fisher
information metric (very similar simulation results were obtained with the
Lebesgue measure).

Given more than one random sample from the uniform, the unrestricted
Bayesian distribution estimator (thick line) always performed better
than the other estimators (as it should by design).  Of course,
asymptotically as $n \rightarrow \infty$, all of the estimates will
converge to the true value. For $n=1$, the Bayesian parameter
estimate performs better; we believe this is due to the (in this
case correct) bias of the prior used for the Bayesian parameter
estimate. The dotted line rises at $n=155$ because the Bayesian
parameter estimate was uncomputable for more than $155$ data samples
(we used Matlab v. 14 to evaluate (\ref{eqn:topEstimate}), and for
$155$ data samples or more the numerator and denominator of
(\ref{eqn:topEstimate}) were determined to be $0$, leading to an
indeterminate estimate).

Three interesting conclusions are supported by the simulation
results. First, the Bayesian estimates do improve significantly over
the maximum likelihood estimate (dashed line). Second, although the
truth is uniform, the unrestricted Bayesian distribution estimate
chooses a non-uniform solution (thick line), which does
significantly better than either of the Bayesian uniform estimates
(thin line and dotted line).  Third, the Bayesian parameter estimate
(dotted line) and the Bayesian uniform distribution estimate (thin
line) perform quite similarly. For $n < 10$, the Bayesian parameter
estimate works better, but for $n
>10$, the Bayesian uniform distribution estimate is slightly better.
Although these two estimates perform similarly, the Bayesian uniform
distribution estimate $[0, 2^{1/n}X_{\max}]$ is a more elegant
solution than the parameter estimate (\ref{eqn:topEstimate}), and is
easier to compute and to work with analytically.

\section{Further Discussion and Open Questions}
We have defined a general Bregman divergence for
functions and distributions that can provide a foundation for results in
statistics, information theory and signal processing. Theorem II.1
is important for these fields because it ties Bregman divergences
to expectation. As shown in Section \ref{sec:bayesian}, Theorem II.1 can be directly
applied to distributions to show that Bayesian distribution estimation simplifies to
expectation when the risk function is a Bregman divergence and the minimizing
distribution is unrestricted.

It is common in Bayesian estimation to interpret the prior as
representing some actual prior knowledge, but in fact prior
knowledge often is not available or is difficult to quantify.
Another approach is to use a prior to capture coarse information
from the data that may be used to stabilize the estimation
\cite{SrivastavaGupta:06, BDA7}. In practice, priors are sometimes
chosen in Bayesian estimation to tame the tail of likelihood
distributions so that expectations will exist when they might
otherwise be infinite \cite{Lehmann:98}. This mathematically
convenient use of priors adds estimation bias that may be
unwarranted by prior knowledge. An alternative to mathematically
convenient priors is to formulate the estimation problem as a
minimization of an expected Bregman divergence between the unknown
distribution and the estimated distribution, and restrict the set of
distributions that can be the minimizer to be a set for which the
Bayesian integral exist. Open questions are how such restrictions
affect the estimation bias and variance, and how to find or define a
``best'' restricted set of distributions for this estimation
approach.

Finally, there are some results for the standard vector Bregman
divergence that have not been extended here. It has been shown that
a standard vector Bregman divergence must be the risk function in
order for the mean to be the minimizer of an expected risk
\cite[Theorems 3 and 4]{Banerjee:05}. The proof of that result
relies heavily on the discrete nature of the underlying vectors, and
it remains an open question as to whether a similar result holds for
the functional Bregman divergence. Another result that has been
shown for the vector case but remains an open question in the
functional case is convergence in probability \cite[Theorem
2]{Banerjee:05}.

\section*{Acknowledgments}
This work was funded in part by the Office of Naval Research, Code
321, Grant \# N00014-05-1-0843. The authors thank Inderjit Dhillon,
Castedo Ellerman, and Galen Shorack for helpful discussions.

\section*{Appendix A: Relevant Definitions and Results from Functional
  Analysis}
This appendix explains the basic definitions
 and results from functional analysis used in this paper. This material
can also be found in standard books on the calculus of variations,
such as the text by Gelfand and Fomin \cite{Gelfand00}.

Let $\left(\Real^d,\Omega,\nu\right)$ be a measure space, where $\nu$ is a
Borel measure $d$ is a positive integer, and define a set of
functions $\mathcal{A} = \left\{a\in L^p(\nu)\st a:\Real^d \to \Real,\;
  a\geq 0 \right\}$ where $1 \leq p \leq \infty$. The subset $\mathcal{A}$ is a convex
subset of $L^p(\nu)$ because for $a_1$, $a_2 \in \mathcal{A}$ and
$ 0 \leq \omega \leq 1$, $\omega a_1 + (1-\omega)a_2 \in
\mathcal{A}$.

\noindent{\bf Definition of continuous linear functionals} \\
The functional $\psi: L^p(\nu)\to \Real$ is linear and continuous if
\begin{enumerate}
\item $\psi[\omega a_1+a_2] = \omega \psi[a_1]+\psi[a_2]$ for any
  $a_1,a_2 \in L^p(\nu)$ and any real number $\omega$;
  and
\item there is a constant C such that $\abs{\psi[a]}\leq C\norm{a}$
  for all $a \in L^p(\nu)$.
\end{enumerate}

\noindent{\bf Functional Derivatives}
\begin{enumerate}
\item Let $\phi$ be a real functional over the normed space
  $L^p(\nu)$. The bounded linear functional $\delta
  \phi[f;\cdot]$ is the Fr\'{e}chet derivative of $\phi$ at $f\in
  L^p(\nu)$ if
\begin{align}
\phi[f+a] - \phi[f] &= \triangle \phi[f;a] \nonumber \\
&= \delta \phi[f;a] + \epsilon[f,a]\left\|a\right\|_{L^p(\nu)}
\label{eqn:FirstDifferential}
\end{align}
\noindent  for all $a\in L^p(\nu)$, with $\epsilon[f,a] \rightarrow 0$
as $\left\|a\right\|_{L^p(\nu)} \rightarrow 0$.
\item When the second variation $\delta^2 \phi$ and the third
  variation $\delta^3 \phi $ exist, they are described by
\begin{eqnarray}
\triangle \phi[f;a] &=& \delta \phi[f;a] +
\frac{1}{2}\delta^2\phi[f;a,a]  \nonumber \\
&& \mbox{} + \epsilon[f,a]\left\|a\right\|^2_{L^p(\nu)}
\label{eqn:SecondDifferential}\\
&=& \delta
\phi[f;a] + \frac{1}{2}\delta^2\phi[f;a,a] \nonumber \\
&& \mbox{} + \frac{1}{6} \delta^3 \phi[f;a,a,a] \nonumber \\
&& \mbox{} + \epsilon[f,a]\left\|a\right\|^3_{L^p(\nu)} ,\nonumber
\end{eqnarray}
where $\epsilon [f,a] \rightarrow 0$ as $\left\|a\right\|_{L^p(\nu)}
\rightarrow 0$. The term $\delta^2\phi[f;a,b]$ is bilinear with
respect to arguments $a$ and $b$, and $\delta^3\phi[f;a,b,c]$ is
trilinear with respect to $a, b$, and $c$.
\item Suppose $ \{ a_n\}, \{f_n\} \subset L^p(\nu)$, moreover $a_n
\rightarrow a$, $f_n \rightarrow f$, where $a,f \in L^p(\nu)$. If
$\phi \in \mathcal{C}^3(L^{p}(\nu);\Real)$ and  $\delta \phi[f;a]$,
$\delta^2 \phi[f;a,a]$, and $\delta^3[f;a,a,a]$ are defined as
above, then $\delta \phi[f_n;a_n] \rightarrow \delta \phi[f;a]$,
$\delta^2\phi[f_n;a_n, a_n] \rightarrow \delta^2 \phi[f;a, a]$, and
$\delta^3\phi[f_n;a_n,a_n,a_n] \rightarrow \delta^3 \phi[f;a,a,a]$,
respectively.
\item The quadratic functional $\delta^2 \phi[f;a,a]$
defined on normed linear space $L^p(\nu)$ is
\emph{\textbf{strongly positive}} if there exists a constant $k>0$
such that $\delta^2 \phi[f;a,a] \geq k
\left\|a\right\|_{L^p(\nu)}^2$ for all $a \in \mathcal{A}$. In a
finite-dimensional space, strong positivity of a quadratic form is
equivalent to the quadratic form being positive definite.
\item From (\ref{eqn:SecondDifferential}),
\begin{eqnarray*}
    \phi[f+a]
    &=&\phi[f]+\delta\phi[f;a]+\frac{1}{2}\delta^2\phi[f;a,a] \\
    &&+ o(\norm{a}^2),\\
    \phi[f] &=&\phi[f+a]-\delta\phi[f+a;a]+ \\
    &&\frac{1}{2}\delta^2\phi[f+a;a,a]+o(\norm{a}^2),
\end{eqnarray*}
where $o(\norm{a}^2)$ stands for a function that goes to zero as
$\norm{a}$ goes to zero, even if it is divided by $\norm{a}^2$.
\noindent Adding the above two equations yields
\begin{eqnarray*}
  0&=&\delta\phi[f;a]-\delta\phi[f+a;a]+\frac{1}{2}\delta^2\phi[f;a,a]
  \\
  &&+\frac{1}{2}\delta^2\phi[f+a;a,a]+o(\norm{a}{}^2),
\end{eqnarray*}
\noindent which is equivalent to
\begin{eqnarray}
\delta\phi[f+a;a]-\delta\phi[f;a]=\delta^2\phi[f;a,a]+o(\norm{a}{}^2),
\label{eqn:ChangeInFirstDifferential}
\end{eqnarray}
because
\begin{multline*}
\abs{\delta^2\phi[f+a;a,a]-\delta^2\phi[f;a,a]}\\
 \leq \norm{\delta^2\phi[f+a;\cdot,\cdot]-
\delta^2\phi[f;\cdot,\cdot]}\norm{a}^2,
\end{multline*}
and we assumed
$\phi\in\mathcal{C}^2$, so $\delta^2\phi[f+a;a,a]-\delta^2\phi[f;a,a]$
is of order $o(\norm{a}^2)$. This shows that the variation of the first
variation of $\phi$ is the second variation of $\phi$. A procedure
like the above can be used to prove that analogous statements hold for higher
variations if they exist.
\end{enumerate}

\noindent \textbf{Functional Optimality Conditions} For a functional
$J$ to have an extremum (minimum) at $f=\hat{f}$, it is necessary
that
\begin{equation*}
\delta J[f;a] = 0 \;\; \mbox{and} \;\; \delta^2J[f;a,a] \geq 0,
\end{equation*}
for  $f=\hat{f}$ and for all admissible functions $a \in
\mathcal{A}$. A sufficient condition for a functional $J[f]$ to have
a minimum for $f=\hat{f}$ is that the first variation $\delta
J[f;a]$ must vanish for $f=\hat{f}$, and its second variation
$\delta^2 J[f;a,a] $ must be strongly positive for $f=\hat{f}$.

\section*{Appendix B: Properties of the Functional Bregman Divergence}
\noindent The Bregman divergence for random variables has some
well-known properties, as reviewed in \cite[Appendix A]{Ghosh:05}.
Here, we establish that the same properties hold for the
functional Bregman divergence (\ref{eqn:BregmanDef}). \\
\noindent{\bf 1. Non-negativity} \\
The functional Bregman divergence is non-negative. \noindent To show
this, define $\tilde\phi:\Real \to \Real$ by
$\tilde\phi(t)=\phi\left[tf+(1-t)g\right]$, $f,g\in\mathcal{A}$. From
the definition of the Fr\'echet derivative,
\begin{equation}\label{eqn:booboo}
    \ddt{\tilde\phi}=\delta\phi[tf+(1-t)g;f-g].
\end{equation}
The function $\tilde\phi$ is convex because $\phi$ is convex by
definition. Then from the mean value theorem there is some $0 \leq
t_0 \leq 1$ such that
\begin{equation}\label{eqn:yogi}
  \tilde\phi(1)-\tilde\phi(0)=
  \ddt{\tilde\phi}(t_0)\geq\ddt{\tilde\phi}(0).
\end{equation}
Because $\tilde\phi(1) = \phi[f]$, $\tilde\phi(0) = \phi[g]$, and
(\ref{eqn:booboo}), subtracting the right-hand side of
(\ref{eqn:yogi}) implies that
\begin{equation}\label{eqn:doghair}
\phi[f]-\phi[g]- \delta\phi[g,f-g] \geq 0.
\end{equation}
If $f=g$, then (\ref{eqn:doghair}) holds in equality. To finish, we
prove the converse. Suppose (\ref{eqn:doghair}) holds in equality;
then
\begin{equation}\label{eqn:last}
\tilde\phi(1) - \tilde\phi(0) = \ddt{\tilde\phi}(0).
\end{equation}
The equation of the straight line connecting $\tilde\phi(0)$ to
$\tilde\phi(1)$ is $\ell(t) = \tilde\phi(0) +
(\tilde\phi(1)-\tilde\phi(0))t$, and the tangent line to the curve
$\tilde\phi$ at $\tilde\phi(0)$ is $y(t) = \tilde\phi(0) +
t\ddt{\tilde\phi}(0)$. Because $\tilde\phi(\tau) = \tilde\phi(0) +
\int_{0}^{\tau} \ddt{\tilde\phi}(t)dt$ and $\ddt{\tilde\phi}(t) \geq
\ddt{\tilde\phi}(0)$ as a direct consequence of convexity, it must
be that $\tilde\phi(t) \geq y(t)$. Convexity also implies that
$\ell(t) \geq \tilde\phi(t)$. However, the assumption that
(\ref{eqn:doghair}) holds in equality implies (\ref{eqn:last}),
which means that $y(t) = \ell(t)$, and thus $\tilde\phi(t) =
\ell(t)$, which is not strictly convex. Because $\phi$ is by
definition strictly convex, it must be true that $\phi[tf + (1-t)g]
< t\phi[f] + (1-t)\phi[g]$ unless $f=g$.  Thus, under the assumption
of equality of (\ref{eqn:doghair}), it must be true that $f=g$.

\noindent{\bf 2. Convexity} \\
The Bregman divergence $d_\phi[f,g]$ is always convex with respect
to $f$. Consider
\begin{eqnarray*}
\triangle d_{\phi}[f,g;a]&=&d_{\phi}[f+a,g] - d_{\phi}[f,g]  \\
&=& \phi[f+a] - \phi[f] - \delta \phi[g; f-g+a] +  \\
&& \delta \phi[g;f-g].
\end{eqnarray*}
\noindent Using linearity in the third term,
\begin{eqnarray*}
\lefteqn{\triangle d_{\phi}[f,g;a]}  \\
&=& \phi[f+a]-\phi[f] -\delta \phi[g;f-g]-  \delta \phi[g;a] \\
&&+ \delta \phi[g;f-g], \\
&=& \phi[f+a] -\phi[f]- \delta \phi[g;a] , \\
&\stackrel{(a)}{=}& \delta \phi[f;a] + \frac{1}{2}\delta^2
\phi[f;a,a] + \epsilon[f,a]\left\|a\right\|^2_{L(\nu)} - \delta
\phi[g;a]\\
&\Rightarrow& \delta^2 d_{\phi}[f,g;a,a] =
\frac{1}{2}\delta^2\phi[f;a,a]
> 0,
\end{eqnarray*}
\noindent where (a) and the conclusion follows from
(\ref{eqn:SecondDifferential}). \\
\noindent{\bf 3. Linearity} \\
The functional Bregman divergence is
linear in the sense that
\begin{eqnarray*}
\lefteqn{d_{(c_1\phi_1+c_2\phi_2)}[f,g]}  \\
&=& (c_1\phi_1+c_2\phi_2)[f]- (c_1\phi_1+c_2\phi_2)[g] -
\\
&&\delta(c_1\phi_1+c_2\phi_2)[g;f-g],  \\
&=& c_1d_{\phi_1}[f,g] + c_2d_{\phi_2}[f,g].
\end{eqnarray*}

\noindent{\bf 4. Equivalence Classes} \\
Partition the set of strictly convex, differentiable functions
$\left\{ \phi \right\}$ on $\mathcal{A}$ into classes with respect
to functional Bregman divergence, so that $\phi_1$ and $\phi_2$
belong to the same class if $d_{\phi_1}[f,g]=d_{\phi_2}[f,g]$ for
all $f,g \in \mathcal{A}$. For brevity we will denote
$d_{\phi_1}[f,g]$ simply by $d_{\phi_1}$. Let $\phi_1 \sim \phi_2$
denote that $\phi_1$ and $\phi_2$ belong to the same class, then
$\sim$ is an equivalence relation because it satisfies the
properties of \emph{reflexivity} (because $d_{\phi_1}=d_{\phi_1}$),
\emph{symmetry} (because if $d_{\phi_1}=d_{\phi_2}$, then
$d_{\phi_2}=d_{\phi_1}$), and \emph{transitivity} (because if
$d_{\phi_1}=d_{\phi_2}$ and $d_{\phi_2}=d_{\phi_3}$, then
$d_{\phi_1}=d_{\phi_3}$).

Further, if $\phi_1 \sim \phi_2$, then they differ only by an affine
transformation. To see this, note that, by assumption, $\phi_1[f]$
$- \phi_1[g]$ $- \delta\phi_1[g;f-g]$ $= \phi_2[f]$$-\phi_2[g]$
$-\delta\phi_2[g;f-g]$, and fix $g$ so $\phi_1[g]$ and $\phi_2[g]$
are constants. By the linearity property, $\delta\phi[g; f-g] =
\delta\phi[g;f] - \delta\phi[g;g]$, and because $g$ is fixed, this
equals $\delta\phi[g;f] + c_0$ where $c_0$ is a scalar constant.
Then $\phi_2[f] = \phi_1[f]  + (\delta\phi_2[g;f] -
\delta\phi_1[g;f]) + c_1$, where $c_1$ is a constant. Thus,
\begin{equation*}
\phi_2[f] = \phi_1[f] + Af +c_1,
\end{equation*}
where $A = \delta\phi_2[g;\cdot] - \delta\phi_1[g;\cdot]$, and thus
$A:\mathcal{A} \rightarrow \Real$ is a linear operator that does not
depend on $f$.

\noindent{\bf 5. Linear Separation} \\
\noindent Fix two non-equal functions $g_1,g_2\in\mathcal{A}$, and
consider the set of all functions in $\mathcal{A}$ that are
equidistant  in terms of functional Bregman divergence from $g_1$
and $g_2$:
\begin{eqnarray*}
\lefteqn{d_{\phi}[f,g_1] = d_{\phi}[f,g_2]} \\
&\Rightarrow& -\phi[g_1]-\delta \phi[g_1;f-g_1] = -\phi[g_2]-\delta
\phi[g_2;f-g_2] \\
&\Rightarrow& -\delta \phi[g_1;f-g_1] = \phi[g_1]-\phi[g_2]-\delta
\phi[g_2;f-g_2].
\end{eqnarray*}
Using linearity the above relationship can be equivalently expressed
as
\begin{eqnarray*}
-\delta \phi[g_1;f] + \delta \phi[g_1;g_1] &=& \phi[g_1] - \phi[g_2]
-\delta \phi[g_2;f] +  \\
&& \delta \phi[g_2;g_2],  \\
\delta \phi[g_2;f] - \delta \phi[g_1;f] &=& \phi[g_1] - \phi[g_2]
-\delta \phi[g_1;g_1] +  \\
&& \delta \phi[g_2;g_2].\\
Lf &=& c,
\end{eqnarray*}
where $L$ is the bounded linear functional defined by $Lf=\delta
\phi[g_2;f] - \delta \phi[g_1;f]$, and $c$ is the constant
corresponding to the right-hand side. In other words, $f$ has to be
in the set $\{a\in\mathcal{A}:La=c\}$, where $c$ is a constant. This
set is a hyperplane.

\noindent{\bf 6. Dual Divergence} \\
Given a pair $(g, \phi)$ where $g \in L^p(\nu)$ and $\phi$ is a
strictly convex twice-continuously Fr\'{e}chet-differentiable
functional, then the function-functional pair $(G, \psi)$ is the
Legendre transform of $(g, \phi)$ \cite{Gelfand00}, if
\begin{eqnarray}
\phi[g] & = & -\psi[G] + \int g(x)G(x)d\nu(x), \label{eqn:111}\\
\delta\phi[g;a] & =& \int G(x)a(x) d\nu(x), \label{eqn:222}
\end{eqnarray}
where $\psi$ is a strictly convex twice-continuously
Fr\'{e}chet-differentiable functional, and $G \in L^q(\nu)$, where
$\frac{1}{p} + \frac{1}{q} = 1$.

Given Legendre transformation pairs $f,g \in L^p(\nu)$ and $F,G \in
L^q(\nu)$,
\begin{equation*}
d_{\phi}[f,g] = d_{\psi}[G,F].
\end{equation*}
The proof begins by substituting (\ref{eqn:111}) and (\ref{eqn:222})
into (\ref{eqn:BregmanDef}):
\begin{eqnarray}
d_{\phi}[f,g] &=& \phi[f] +\psi[G] - \int g(x)G(x) d\nu(x) \nonumber \\
&-&\int G(x)(f-g)(x) d\nu(x) \nonumber\\
&=& \phi[f] + \psi[G] - \int G(x)f(x) d\nu(x). \label{eqn:555}
\end{eqnarray}
Applying the Legendre transformation to $(G,\psi)$ implies that
\begin{eqnarray}
\psi[G] & = & -\phi[g] + \int g(x)G(x)d\nu(x) \label{eqn:333}\\
\delta\psi[G;a] & =& \int g(x)a(x) d\nu(x) \label{eqn:444}.
\end{eqnarray}
Using (\ref{eqn:333}) and (\ref{eqn:444}), $d_{\psi}[G,F]$ can be
reduced to (\ref{eqn:555}).

\noindent{\bf 7. Generalized Pythagorean Inequality}

\noindent For any $f,g,h \in \mathcal{A}$,
\begin{equation*}
d_{\phi}[f,h] = d_{\phi}[f,g]+d_{\phi}[g,h] + \delta \phi[g;f-g] -
\delta \phi[h;f-g].
\end{equation*}
This can be derived as follows:
\begin{eqnarray*}
\lefteqn{d_{\phi}[f,g]+d_{\phi}[g,h]} \\
&=& \phi[f] - \phi[h] - \delta \phi[g;f-g] - \delta \phi[h;g-h] \\
&=& \phi[f] - \phi[h] - \delta \phi[h;f-h] + \delta \phi[h;f-h]\\
&& -\delta \phi[g;f-g] - \delta \phi[h;g-h]\\
&=& d_{\phi}[f,h] + \delta \phi[h;f-g] - \delta \phi[g;f-g],
\end{eqnarray*}
where the last line follows from the definition of the functional
Bregman divergence and the linearity of the fourth and last terms.

\section*{Appendix C: Proofs}
\subsection{Proof of Proposition I.2}
We give a constructive proof that there is a corresponding
functional Bregman divergence $d_{\phi}[f,g]$ for a specific choice
of $\phi:\mathcal{A}^1 \to\Real$, where $\mathcal{A}^1$ is the set
of functions $\mathcal{A}$ with $p=1$, and where $\nu=\sum_{i=1}^n
\delta_{c_i}$ and $f,g \in \mathcal{A}^1$. Here, $\delta_{x}$ is the
Dirac measure (such that all mass is concentrated at $x$) and
$\{c_1, c_2, \ldots, c_n\}$ is a collection of $n$ distinct points
in $\Real^d$.

For any $x \in \Real^n$, define $\phi[f] = \tilde\phi(x_1, x_2,
\ldots, x_n)$, where $f(c_1) = x_1, f(c_2) =x_2, \ldots, f(c_n) =
x_n$. Then the difference is
\begin{eqnarray*}
\lefteqn{\Delta\phi[f;a] = \phi[f+a]-\phi[f]} \\
 &=&\tilde\phi\left((f+a)(c_1),\ldots,(f+a)(c_n)\right)-
  \tilde\phi\left(x_1,\ldots,x_n\right)\\
  &=&\tilde\phi\left(x_1+a(c_1),\ldots,x_n+a(c_n)\right)-
  \tilde\phi\left(x_1,\ldots,x_n\right).
\end{eqnarray*}
Let $a_i$ be short hand for $a(c_i)$, and use the Taylor expansion
for functions of several variables to yield
\begin{equation*}
\Delta \phi[f;a]=\nabla \tilde\phi(x_1,\ldots,x_n)^T
  (a_1,\ldots,a_n)+\epsilon[f,a]\norm{a}_{L^1}.
\end{equation*}
Therefore,
\begin{equation*}
\delta \phi[f;a] = \nabla\tilde\phi(x_1,\ldots,x_n)^T
(a_1,\ldots,a_n) = \nabla \tilde\phi(x)^T a,
\end{equation*}
where $x = (x_1,x_2, \ldots,x_n)$ and $a = (a_1,\ldots,a_n)$. Thus,
from (3), the functional Bregman divergence definition
(\ref{eqn:BregmanDef}) for $\phi$ is equivalent to the standard
vector Bregman divergence:
\begin{eqnarray}
d_{\tilde{\phi}}[f,g] &=& \phi[f] - \phi[g] - \delta
\phi[g;f-g] \nonumber \\
&=& \tilde\phi(x) - \tilde\phi(y) - \nabla \tilde\phi(y)^T(x-y).
\label{eqn:discreteBregmanDiv}
\end{eqnarray}

\subsection{Proof of Proposition I.3}
First, we give a constructive proof of the first part of the
proposition by showing that given a $B_{s, \nu}$, there is an
equivalent functional divergence $d_{\phi}$. Then, the second part
of the proposition is shown by example: we prove that the squared
bias functional Bregman divergence given in Section
\ref{sec:biasExample} is a functional Bregman divergence that cannot
be defined as a pointwise Bregman divergence.

Note that the integral to calculate $B_{s, \nu}$ is not always
finite. To ensure finite $B_{s, \nu}$, we explicitly constrain
$\lim_{x\to 0} s'(x)$ and $\lim_{x \rightarrow 0} s(x)$ to be
finite. From the assumption that $s$ is strictly convex, $s$ must be
continuous on $(0, \infty)$. Recall from the assumptions that the
measure $\nu$ is finite, and that the function $s$ is differentiable
on $(0, \infty)$.

Given a $B_{s,\nu}$, define the continuously differentiable function
\begin{equation*}
    \tilde{s}(x)=
    \begin{cases}
      s(x) &x\geq 0\\
      -s(-x)+2s(0) & x<0.
    \end{cases}
\end{equation*}
Specify $\phi: L^\infty(\nu) \to \Real$ as
\begin{equation*}
\phi[f]=\int_X \tilde{s}(f(x))d\nu.
\end{equation*}
Note that if $f \geq 0$,
\begin{equation*}
  \phi[f]=\int_Xs(f(x))d\nu.
\end{equation*}
Because $\tilde{s}$ is continuous on $\Real$, $\tilde{s}(f) \in
L^{\infty}$ whenever $f \in L^{\infty}$, the integrals always make
sense.

It remains to be shown that $\delta \phi [f;\cdot]$ completes the
equivalence when $f \geq 0 $. For $h \in L^{\infty}$,
\begin{align*}
  \phi[f+h]-\phi[f] &=\int_X \tilde{s}(f(x)+h(x))d\nu- \int_X
  s(f(x))d\nu\\
  &=\int_X \tilde{s}(f(x)+h(x))-s(f(x))d\nu\\
  &=\int_X \tilde{s}'(f(x))h(x)+ \epsilon\left(f(x),h(x)\right)h(x)
    d\nu \\
  &=\int_X s'(f(x))h(x)+ \epsilon\left(f(x),h(x)\right)h(x) d\nu,
\end{align*}
where we used the fact that
\begin{eqnarray*}
\lefteqn{\tilde{s}(f(x)+h(x))}\\
& =&\tilde{s}(f(x))+\left(\tilde{s}'(f(x))+\epsilon(f(x),h(x))\right)h(x)\\
 & =&s(f(x))+\left(s'(f(x))+\epsilon(f(x),h(x))\right)h(x),
\end{eqnarray*}
because $f\geq 0$. On the other hand, if $h(x)=0$ then
$\epsilon(f(x),h(x))=0$, and if $h(x)\neq 0$ then
\begin{equation*}
  \abs{\epsilon(f(x),h(x))}\leq
  \abs{\frac{\tilde{s}(f(x)+h(x))-\tilde{s}(f(x))}{h(x)}}+\abs{s'(f(x))}.
\end{equation*}

Suppose $\{h_n\}\subset L^\infty(\nu)$ such that $h_n\to 0$. Then
there is a measurable set $E$ such that its complement is of measure
$0$ and $h_n\to 0$ uniformly on $E$. There is some $N>0$ such that
for any $n>N$, $\abs{h_n(x)}\leq \epsilon$ for all $x\in E$. Without
loss of generality, assume that there is some $M>0$ such that for
all $x\in E$, $\abs{f(x)}\leq M$. Since $\tilde{s}$ is continuously
differentiable, there is a $K>0$ such that $\max\{\tilde{s}'(t)\st
  t\in[-M-\epsilon,M+\epsilon]\}\leq K$, and by the mean value theorem
  \begin{equation*}
    \abs{\frac{\tilde{s}(f(x)+h(x))-\tilde{s}(f(x))}{h(x)}} \leq K,
  \end{equation*}
for almost all $x\in X$. Then
\begin{equation*}
  \abs{\epsilon(f(x),h(x))}\leq 2K,
\end{equation*}
except on a set of measure $0$. The fact that $h(x)\to 0$ almost
everywhere implies that $\abs{\epsilon(f(x),h(x))}\to 0$ almost
everywhere, and by Lebesgue's dominated convergence theorem, the
corresponding integral goes to $0$. As a result, the Fr\'echet
derivative of $\phi$ is
\begin{equation}\label{eqn:daisy}
  \delta\phi[f;h]=\int_X s'(f(x))h(x)d\nu.
\end{equation}
Thus the functional Bregman divergence is equivalent to the given
pointwise $B_{s,\nu}$.

We additionally note that the assumptions that $f \in L^{\infty}$
and that the measure $\nu$ is finite are necessary for this proof.
Counterexamples can be constructed if $f \in L^p$ or $\nu(X) =
\infty$ such that the Fr\'echet derivative of $\phi$ does not obey
(\ref{eqn:daisy}). This concludes the first part of the proof.

To show that the squared bias functional Bregman divergence given in
Section \ref{sec:biasExample} is an example of a functional Bregman
divergence that cannot be defined as a pointwise Bregman divergence
we prove that the converse statement leads to a contradiction.

Suppose $(X,\Sigma,\nu)$ and $(X,\Sigma,\mu)$ are measure spaces where
$\nu$ is a non-zero $\sigma$-finite measure and that there is a
differentiable function $f:(0,\infty)\to\R$ such that
\begin{equation}
  \label{eqn:whale}
  \left(\int \xi d\nu\right)^2=\int f(\xi) d\mu,
\end{equation}
where $\xi\in \mathcal{A}^1$, the set of functions $\mathcal{A}$ with
$p=1$. Let $f(0)=\lim_{x\to 0}f(x)$, which can be
finite or infinite, and let $\alpha$ be any real number. Then
\begin{align*}
  \int f(\alpha\xi)d\mu =\left(\int \alpha\xi d\nu\right)^2
  &=\alpha^2\left(\int \xi  d\nu\right)^2\\
  &=\alpha^2\int  f(\xi) d\mu.
\end{align*}
Because $\nu$ is $\sigma$-finite, there is a measurable set $E$ such
that $0<\abs{\nu(E)}<\infty$. Let $X\backslash E$ denote the
complement of $E$ in $X$. Then
\begin{align*}
  \alpha^2\nu^2(E)  &=\alpha^2\left(\int I_E d\nu\right)^2\\
  &=\alpha^2\int  f(I_E) d\mu\\
  &=\alpha^2\int_{X\backslash E}f(0) d\mu+ \alpha^2\int_E  f(1)
  d\mu\\
  &=\alpha^2f(0)\mu(X\backslash E)+\alpha^2f(1)\mu(E).
\end{align*}
Also,
\begin{equation*}
\alpha^2\nu^2(E) = \left(\int \alpha I_E d\nu\right)^2.
\end{equation*}
However,
\begin{align*}
  \left(\int \alpha I_E d\nu\right)^2 &= \int f(\alpha I_E)d\mu\\
  &=\int_{X\backslash E}f(\alpha I_E) d\mu+
  \int_Ef(\alpha I_E) d\mu\\
  &=f(0)\mu(X\backslash E)+f(\alpha)\mu(E);
\end{align*}
so one can conclude that
\begin{eqnarray}\label{eqn:dolphin}
\lefteqn{\alpha^2f(0)\mu(X\backslash E)+
  \alpha^2f(1)\mu(E)}\nonumber\\
  &=&f(0)\mu(X\backslash E)+f(\alpha)\mu(E).
\end{eqnarray}

Apply equation (\ref{eqn:whale}) for $\xi=0$ to yield
\begin{equation*}
  0=\left(\int 0 d\nu\right)^2=\int f(0)d\mu=f(0)\mu(X).
\end{equation*}
Since $\abs{\nu(E)}>0$, $\mu(X)\neq 0$, so it must be that $f(0)=0$,
and (\ref{eqn:dolphin}) becomes
\begin{equation*}
  \alpha^2\nu^2(E)=\alpha^2f(1)\mu(E)=f(\alpha)\mu(E)\quad
  \forall\alpha\in\R.
\end{equation*}

The first equation implies that $\mu(E)\neq 0$. The second equation
determines the function $f$ completely:
\begin{equation*}
  f(\alpha)=f(1)\alpha^2.
\end{equation*}
Then (\ref{eqn:whale}) becomes
\begin{equation*}
  \left(\int \xi d\nu\right)^2=\int f(1)\xi^2 d\mu.
\end{equation*}

Consider any two disjoint measurable sets, $E_1$ and $E_2$, with
finite nonzero measure. Define $\xi_1 = I_{E_1}$ and $\xi_2 =
I_{E_2}$. Then $\xi=\xi_1+\xi_2$ and $\xi_1\xi_2= I_{E_1}I_{E_2}=0$.
Equation (\ref{eqn:whale}) becomes
\begin{equation}
  \int \xi_1d\nu\int \xi_2 d\nu=f(1) \int \xi_1\xi_2 d\mu.
\end{equation}
This implies the following contradiction:
\begin{equation}
  \int \xi_1d\nu\int \xi_2 d\nu=\nu(E_1)\nu(E_2)\neq 0,
\end{equation}
but
\begin{equation}
  f(1) \int \xi_1\xi_2 d\mu=0.
\end{equation}

\subsection{Proof of Theorem II.1} Let
\begin{eqnarray}
J[g] &=& E_{P_F}[d_{\phi}(F,g)] = \int_{M} d_{\phi}[f,g]P(f)dM \nonumber \\
&=& \int_{M} (\phi[f] - \phi[g] - \delta
\phi[g;f-g])P(f)dM,\label{eqn:BregmanObjective}
\end{eqnarray}
\noindent where (\ref{eqn:BregmanObjective}) follows by substituting
the definition of Bregman divergence (\ref{eqn:BregmanDef}).
Consider the increment
\begin{align}
  \Delta J[g;a] &= J[g+a]-J[g] \label{eqn:Increment} \\
  &= -\int_M \left(\phi[g+a] - \phi[g] \right)P(f)dM  \nonumber\\
  &\quad- \int_M \left(\delta \phi[g+a; f -g -a]\right.\nonumber\\
  &\quad\left.- \delta \phi[g;f-g]\right) P(f)dM, \label{eqn:BregmanIncrement}
\end{align}
where (\ref{eqn:BregmanIncrement}) follows from substituting
(\ref{eqn:BregmanObjective}) into (\ref{eqn:Increment}). Using the
definition of the differential of a functional (see Appendix A,
(\ref{eqn:FirstDifferential})), the first integrand in
(\ref{eqn:BregmanIncrement}) can be written as
\begin{eqnarray}
\phi[g+a] - \phi[g] &=& \delta \phi[g;a] + \epsilon[g,a] \left\| a
\right\|_{L^1(\nu)}. \label{eqn:BregmanStep1}
\end{eqnarray}
Take the second integrand of (\ref{eqn:BregmanIncrement}), and
subtract and add $\delta \phi[g;f-g-a]$,
\begin{eqnarray}
\lefteqn{\delta\phi[g+a;f-g-a] - \delta \phi[g;f-g]  \nonumber} \\
&=& \delta \phi[g+a; f-g-a] - \delta \phi[g; f-g-a] \nonumber \\
&& \mbox{} + \delta \phi[g; f-g-a] - \delta \phi[g;f-g] \nonumber
\\
&\stackrel{(a)}{=}& \delta^2 \phi[g;f-g-a,a] + \epsilon[g,a]\left\|
a\right\|_{L^1(\nu)} + \delta \phi[g;f-g] \nonumber \\
&& \mbox{} - \delta \phi[g;a] - \delta\phi[g;f-g] \nonumber \\
&\stackrel{(b)}{=}& \delta^2 \phi[g; f-g,a]- \delta^2 \phi[g;a,a] +
\epsilon[g,a] \left\| a \right\|_{L^1(\nu)} \nonumber \\
&& \mbox{} - \delta \phi[g;a] \label{eqn:BregmanStep2}
\end{eqnarray}
\noindent where $(a)$ follows from
(\ref{eqn:ChangeInFirstDifferential})  and the linearity of the
third term, and $(b)$ follows from the linearity of the first term.
Substitute (\ref{eqn:BregmanStep1}) and (\ref{eqn:BregmanStep2})
into (\ref{eqn:BregmanIncrement}),
\begin{eqnarray*}
\triangle J[g;a] &=& - \int_M \Big( \delta^2 \phi[g;f-g,a] -
\delta^2 \phi[g;a,a] \\
&& \mbox{} + \epsilon[g,a] \left\|a\right\|_{L^1(\nu)} \Big) P(f)
dM.
\end{eqnarray*}
\noindent Note that the term $ \delta^2 \phi[g;a,a]$ is of order
$\left\|a\right\|^2_{L^1(\nu)}$, that is, $\left\|\delta^2
\phi[g;a,a]\right\|_{L^1(\nu)} \leq m\left\| a
\right\|^2_{L^1(\nu)}$ for some constant $m$. Therefore,
\begin{eqnarray*}
 &&\lim_{\left\|a\right\|_{L^1(\nu)} \rightarrow 0}
 \frac{\left\| J[g+ a] -  J[g] -
 \delta J[g;a]
\right\|_{L^1(\nu)}}{\left\|a \right\|_{L^1(\nu)}} = 0,
\end{eqnarray*}
where,
\begin{eqnarray} \delta J[g;a] &=& -\int_M \delta^2
\phi[g;f-g,a]P(f) dM. \label{eqn:VariationOfBregmanObjective}
\end{eqnarray}
\noindent For fixed $a$, $\delta^2 \phi[g; \cdot, a]$ is a bounded
linear functional in the second argument, so the integration and the
functional can be interchanged in
(\ref{eqn:VariationOfBregmanObjective}), which becomes
\begin{equation*}
 \delta J[g;a] = -\delta^2 \phi\left[g; \int_M \left(f-g \right)P(f)dM, a \right].
\end{equation*}
Using the functional optimality conditions (stated in Appendix A),
$J[g]$ has an extremum for $g=\hat{g}$ if
\begin{equation}
\delta^2 \phi\left[\hat{g}; \int_M \left(f-\hat{g} \right)P(f) dM,a
\right] = 0. \label{eqn:BilinearSecondVariation}
\end{equation}
\noindent Set $a = \int_M \left(f-\hat{g} \right)P(f) dM$ in
(\ref{eqn:BilinearSecondVariation}) and use the assumption that the
quadratic functional $\delta^2 \phi[g;a,a]$ is strongly positive,
which implies that the above functional can be zero if and only if
$a=0$, that is,
\begin{eqnarray}
0&=&\int_M (f-\hat{g}) P(f)dM, \label{eqn:FatDog} \\
\hat{g} &=& E_{P_f} [F], \label{eqn:BregmanExtremum}
\end{eqnarray}
where the last line holds if the expectation exists (i.e. if the
measure is well-defined and the expectation is finite). Because a
Bregman divergence is not necessarily convex in its second argument,
it is not yet established that the above unique extremum is a
minimum. To see that (\ref{eqn:BregmanExtremum}) is in fact a
minimum of $J[g]$, from the functional optimality conditions it is
enough to show that $\delta^2 J[\hat{g};a,a]$ is strongly positive.
To show this, for $b \in \mathcal{A}$, consider
\begin{eqnarray}
\lefteqn{\delta J[g+b;a] - \delta J[g;a] \nonumber} \\
&\stackrel{(c)}{=}& -\int_M \big( \delta^2 \phi[g+b;f-g-b,a]
\nonumber \\
&& \mbox{} - \delta^2\phi[g;f-g,a] \big) P(f) dM \nonumber \\
&\stackrel{(d)}{=}& -\int_M \big( \delta^2 \phi[g+b;f-g-b,a] -
\delta^2\phi[g;f-g-b,a] \nonumber \\
&& \mbox{} + \delta^2\phi[g;f-g-b,a] - \delta^2\phi[g;f-g,a]\big)P(f) dM \nonumber \\
&\stackrel{(e)}{=}& -\int_M \big( \delta^3 \phi[g;f-g-b,a,b] +
\epsilon
[g,a,b]\left\|b\right\|_{L^1(\nu)} \nonumber \\
&& \mbox{} + \delta^2 \phi[g;f-g,a] - \delta^2 \phi[g;b,a] \nonumber \\
&& \mbox{} - \delta^2 \phi[g;f-g,a] \big) P(f) dM \nonumber \\
&\stackrel{(f)}{=}& -\int_M \big(\delta^3 \phi[g;f-g,a,b] -\delta^3
\phi[g;b,a,b] \nonumber \\
&& \mbox{} + \epsilon[g,a,b]\left\|b\right\|_{L^1(\nu)} - \delta^2
\phi[g;b,a] \big) P(f) dM, \label{eqn:EquationF}
\end{eqnarray}
\noindent where $(c)$ follows from using integral
(\ref{eqn:VariationOfBregmanObjective}); $(d)$ from subtracting and
adding $\delta^2 \phi[g;f-g-b,a]$; $(e)$ from the fact that the
variation of the second variation of $\phi$ is the third variation
of $\phi$ \cite{Edwards:95}; and $(f)$ from the linearity of the
first term and cancellation of the third and fifth terms. Note that
in (\ref{eqn:EquationF}) for fixed $a$, the term $\delta^3
\phi[g;b,a,b]$ is of order $\left\|b\right\|^2_{L^1(\nu)} $, while
the first and the last terms are of order
$\left\|b\right\|_{L^1(\nu)}$. Therefore,
\begin{equation*}
 \lim_{\left\|b\right\|_{L^1(\nu)} \rightarrow 0}
 \frac{\left\| \delta J[g+b;a] - \delta J[g;a] -
 \delta^2J[g;a,b]
\right\|_{L^1(\nu)}}{\left\|b \right\|_{L^1(\nu)}} = 0,
 \end{equation*}
\noindent where
 \begin{eqnarray}
\delta^2 J[g;a,b] &=& -\int_M \delta^3 \phi[g;f-g,a,b]P(f)dM
\nonumber \\
&+& \int_M \delta^2 \phi[g;a,b] P(f) dM.
\label{eqn:BilinearBregmanFunctional}
\end{eqnarray}
Substitute $b = a$, $g=\hat{g}$ and interchange integration and
the continuous functional $\delta^3\phi$ in the first integral of
(\ref{eqn:BilinearBregmanFunctional}), then
\begin{eqnarray}
\delta^2 J[\hat{g};a,a] &=& -\delta^3 \phi
\left[\hat{g};\int_M(f-\hat{g})
P(f)dM,a,a \right] \nonumber \\
&& \mbox{} + \int_M \delta^2
\phi[\hat{g};a,a] P(f) dM \nonumber \\
 &=& \int_M \delta^2
\phi[\hat{g};a,a] P(f) dM \label{eqn:SecondVariationBregman} \\
&\geq& \int_M k \left\|a\right\|^2_{L^1(\nu)} P(f)dM \nonumber \\
&=& k \left\|a\right\|^2_{L^1(\nu)} \;>\; 0,
\label{eqn:StronglyPositiveBregman}
\end{eqnarray}
\noindent where (\ref{eqn:SecondVariationBregman}) follows from
(\ref{eqn:FatDog}), and (\ref{eqn:StronglyPositiveBregman}) follows
from the strong positivity of $\delta^2 \phi[\hat{g};a,a]$.
\noindent Therefore, from (\ref{eqn:StronglyPositiveBregman}) and
the functional optimality conditions, $\hat{g}$ is the minimum.

\subsection{Derivation of the Bayesian Distribution-based Uniform Estimate Restricted
to a Uniform Minimizer} Let $f(x) = 1/a$ for all $0 \leq x \leq a$
and $g(x) = 1/b$ for all $0 \leq x \leq b$. Assume at first that $b
> a$; then the total squared difference between $f$ and $g$ is
\begin{eqnarray*}
\int_x (f(x) - g(x))^2 dx &=& a\left(\frac{1}{a} -
\frac{1}{b}\right)^2 + (b-a)\left(\frac{1}{b}\right)^2\\
&=&\frac{b-a}{ab}\\
&=&\frac{|b-a|}{ab},
\end{eqnarray*}
where the last line does not require the assumption that $b > a$.

In this case, the integral (\ref{eqn:stravinsky}) is over the
one-dimensional manifold of uniform distributions $\mathcal{U}$; a
Riemannian metric can be formed by using the differential arc
element to convert Lebesgue measure on the set $\mathcal{U}$ to a
measure on the set of parameters $a$ such that
(\ref{eqn:stravinsky}) is re-formulated in terms of the parameters
for ease of calculation:
\begin{equation}\label{eqn:uniform2}
b^* = \arg \min_{b \in \Real^+} \int_{a=X_{\max}}^{\infty}
\frac{|b-a|}{ab} \frac{1}{a^n}\left\|\frac{df}{da}\right\|_2da,
\end{equation}
where $a^n$ is the likelihood of the $n$ data points being drawn
from a uniform distribution $[0, a]$, and the estimated distribution
is uniform on $[0, b^*]$. The differential arc element
$\left\|\frac{df}{da}\right\|_2$ can be calculated by expanding
$df/da$ in terms of the Haar orthonormal basis
$\{\frac{1}{\sqrt{a}}, \phi_{jk}(x)\}$, which forms a complete
orthonormal basis for the interval $0 \leq x \leq a$, and then the
required norm is equivalent to the norm of the basis coefficients of
the orthonormal expansion:
\begin{equation}\label{eqn:lebesgue}
\left\|\frac{df}{da}\right\|_2 = \frac{1}{a^{3/2}}.
\end{equation}

For estimation problems, the measure determined by the Fisher
information metric may be more appropriate than Lebesgue measure
\cite{Kass, Amari:00, Lebanon:05}. Then
\begin{equation}\label{eqn:gofish}
dM = |I(a)|^{\frac{1}{2}} da,
\end{equation}
where $I$ is the Fisher information matrix. For the one-dimensional
manifold $M$ formed by the set of scaled uniform distributions
$\mathcal{U}$, the Fisher information matrix is
\begin{eqnarray*}
I(a) &=& E_X\left[\left(\frac{d \log \frac{1}{a}}{da}\right)^2\right]\\
& = & \int_{0}^{a} \frac{1}{a^2}\frac{1}{a} dx = \frac{1}{a^2},
\end{eqnarray*}
so that the differential element is $dM = \frac{da}{a}$.

We solve (\ref{eqn:stravinsky}) using the Lebesgue measure
(\ref{eqn:lebesgue}); the solution with the Fisher differential
element follows the same logic. Then (\ref{eqn:uniform2}) is
equivalent to
\begin{equation*}
\arg \min_{b} J(b)= \int_{a=X_{\max}}^{\infty} \frac{|b-a|}{ab}
\frac{1}{a^{n+3/2}}da
\end{equation*}
\begin{eqnarray*}
&=&\int_{a=X_{\max}}^{b} \frac{b-a}{ab} \frac{da}{a^{n+3/2}} +
\int_{b}^{\infty}  \frac{a-b}{ab} \frac{da}{a^{n+3/2}}\\
&=& \frac{2}{(n+1/2)(n+3/2)b^{n+3/2}} -
\frac{1}{b(n+\frac{1}{2})X_{\max}^{n+\frac{1}{2}}}  \\
&& \mbox{}+ \frac{1}{(n+3/2)X_{\max}^{n+3/2}} \; .
\end{eqnarray*}
The minimum is found by setting the first derivative to zero:
\begin{eqnarray*}
J'(\hat{b}) &=& \frac{2}{(n+1/2)(n+3/2)}\frac{\left(n+3/2\right)}
{\hat{b}^{n+5/2}} \nonumber\\
&& \mbox{} +
\frac{1}{\hat{b}^2(n+1/2)X_{\max}^{n+1/2}} = 0 \nonumber\\
\Rightarrow \hat{b} &=& 2^{\frac{1}{n+1/2}} X_{\max}.
\end{eqnarray*}
To establish that $\hat{b}$ is in fact a minimum, note that
\begin{equation*}
J''(\hat{b}) = \frac{1}{\hat{b}X_{\max}^{n+1/2}} =
\frac{1}{2^{\frac{3}{n+1/6}}X_{\max}^{n+7/2}} > 0.
\end{equation*}
Thus, the restricted Bayesian estimate is the uniform distribution
over $[0, 2^{\frac{1}{n+1/2}} X_{\max}]$.

\bibliographystyle{IEEEtran}
\bibliography{ieeerefs}
\end{document}